# Approximate Expressions to the Reactive Near-Field Losses of Body-Implanted Antennas

Mingxiang Gao, Zvonimir Šipuš *Senior Member, IEEE*, Anja K. Skrivervik

*Abstract*—**Implantable bioelectronics often relies on an RF-wireless link for communication and/or remote powering. Propagation through biological media is very lossy, and previous work has shown that these losses can be separated into three parts: the losses incurred by the propagating fields, the reflections at media interfaces, and the coupling of the reactive near field and the lossy body. The first two are unavoidable, but a clever antenna design can minimize the near-field losses. A good physical understanding of this particular loss phenomenon is thus very desirable. Unfortunately, previous work does not take the implantation depth into account and is thus valid only for deep implants. In this contribution, we propose approximate expressions to the near field losses of antennas implanted in biological hosts considering the encapsulation size, the frequency, the characteristics of the host medium, and the implantation depth. They are obtained using a spherical wave expansion, and are useful for realistic implantation scenarios. This is demonstrated by first comparing the results obtained using these expressions with rigorous computations for two canonical phantoms: a spherical phantom and a planar phantom. Finally, the usefulness of the obtained expressions is illustrated in the practical realization of a capsule-shaped implanted antenna.**

## I. INTRODUCTION

MINIATURE implanted bioelectronics provide breakthrough capabilities for biomedical research [1]–[3]. Especially in the last decade, with the development of advanced materials and micro-fabrication technologies, miniaturized implanted electronics are used in applications from implantable neural interfaces to cell-level implants. For instance, advanced neuroengineering platforms have developed into long-lived neural interfaces with diverse operational modes, in which closed-loop operation through a wireless link becomes essential [4]. Thus, for most body-implanted devices, wireless communication and power transfer through the lossy body have become an indispensable task [4], [5]. To achieve an efficient wireless link, it becomes crucial to understand the mechanisms of radiation losses of implanted antennas.

The radiation properties of arbitrary antennas in free space have been well investigated [6]–[8]. However, for antennas fully placed in lossy biological tissues, dissipation and scattering in the lossy media become the main cause of radiation losses, and the geometry of the body phantom also greatly influences the antenna characteristics. By looking into body phantoms with simplified geometries, such as spherical or planar body models, the radiation process of an implanted source can be analyzed from a macro perspective [9]–[16]. On this basis, studies have been carried out to explore the physical limitations of implanted antennas and the optimum design of wireless body area networks [12]–[14].

One research goal is to gain insight into the dissipation due to the lossy host body by establishing approximate analytical expressions. In this way, it is possible to quantitatively approximate the losses due to different contributions, which provides understanding and guidelines for a specific design and antenna optimization. In [14], the loss mechanisms for implanted antennas are analyzed, and are divided into three main contributions: the losses due to the dissipation in the reactive near field, the losses owing to the propagating field absorption, and the reflection losses. The antenna designer has little influence on propagating and reflection losses but a careful design can significantly reduce the near-field losses. Thus, the reactive near-field losses require a full understanding and effective quantitative estimation.

Previous work has proposed approximate analytic expressions for these near field losses taking into account key factors like the encapsulation size and the lossy medium into which the antenna is implanted [14], [16], but not the implantation depth. In this work, we propose a modified approach yielding new approximate expressions to assess the reactive near-field losses, which obtain accurate results for both deep and shallow implants, as the implantation depth is accounted for. The validity of the proposed expressions has been demonstrated in the examples of an implanted dipole source within spherical and planar body phantoms. Furthermore, a realistic capsule-shaped antenna implanted within a cube-shaped body model has also been investigated and well demonstrated the utility of approximate expressions in a general implantation site.

## II. MODEL AND METHOD

### A. Spherical Body Model

Let us consider first a simplified canonical model of a biological body with an implanted capsule: an elementary source placed in a spherical body phantom, as depicted in Fig. 1. The implanted antenna, an elementary electric or magnetic

M. Gao and A. K. Skrivervik are with the Microwaves and Antennas Group, Ecole Polytechnique Federale de Lausanne, Lausanne, Switzerland (e-mail: mingxiang.gao@epfl.ch, anja.skrivervik@epfl.ch).

Z. Sipus is with Faculty of Electrical Engineering and Computing, University of Zagreb, Zagreb, Croatia (e-mail: zvonimir.sipus@fer.hr).

dipole, is surrounded by a small air sphere, of radius $r_{\mathrm{impl}}$, which roughly represents the dimension of the antenna encapsulation. The source is implanted at a depth $d$ within the spherical phantom, the latter having variable radius $r_{\mathrm{body}}$. To simulate the lossy biological tissue, the complex permittivity derived from the four-region Cole–Cole model [17] is applied to the medium of body phantom. The phantom is set to be homogeneous to facilitate the initial analysis.

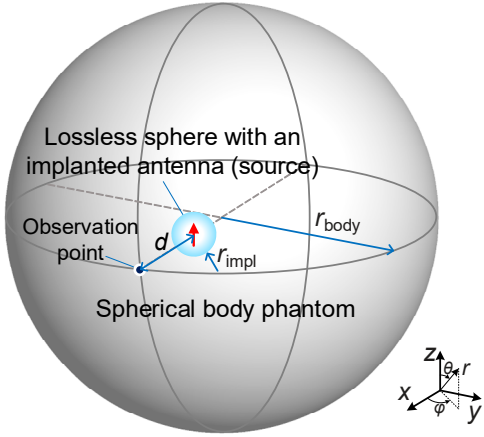

Fig. 1. View of the spherical body model with an elementary dipole source implanted at a depth of $d$.

To characterize the losses of implanted antennas, the radiation efficiency and maximum power density are two key performance indicators (KPIs) of interest [14]. The radiation efficiency provides a comprehensive indicator of the dissipated losses due to the presence of the body phantom. It can be calculated from the ratio of the total radiated power reaching outside of the phantom to the power entering the body medium through the encapsulation around the antenna [13]. The maximum power density reaching the body interface evaluates the dissipated losses on the shortest communication link from the source to the phantom–air interface. For most implantation scenarios, especially shallow implants, the maximum power density is a more practical KPI as it shows the minimum of the dissipated losses, which is related to the upper limit of the radiation efficiency.

### B. Spherical Wave Expansion Method

For the spherical body model shown in Fig. 1, the spherical wave expansion (SWE) method is applied, allowing us to analytically compute the EM field distribution [18]–[20]. The complex EM field quantities $\mathbf{E}$ and $\mathbf{H}$ in regions without presence of any source are expressed using vector spherical harmonics $\mathbf{M}$ and $\mathbf{N}$:

$$\begin{Bmatrix} \mathbf{E} \\ -i\eta\mathbf{H} \end{Bmatrix} = \sum_{n,m} a_{mn} \begin{Bmatrix} \mathbf{M}_{mn} \\ \mathbf{N}_{mn} \end{Bmatrix} + b_{mn} \begin{Bmatrix} \mathbf{N}_{mn} \\ \mathbf{M}_{mn} \end{Bmatrix}, \quad (1)$$

where $\sum_{n,m} = \sum_{n=1}^{+\infty} \sum_{m=-n}^{n}$, $a_{mn}$ and $b_{mn}$ are the spherical modal coefficients, $m$ and $n$ are the mode indexes, $\eta$ is the intrinsic impedance in the medium, and $\mathbf{M}_{mn}$ and $\mathbf{N}_{mn}$ are both functions of the scalar function $\psi_{mn}$. More specifically, here $\psi_{mn} = (1/kr)\hat{Z}_n(kr)P_n^{|m|}(\cos\theta)e^{im\varphi}$, which is the solution of the scalar Helmholtz equation in spherical coordinate. $\hat{Z}_n$ represent the spherical Bessel functions $\hat{B}_n$ or Hankel functions $\hat{H}_n$ used by Schelkunoff [19], and $P_n^m$ are the associated Legendre functions.

With the SWE, the source can be placed at any point inside the spherical body phantom, and the mode matching technique based on spherical wave functions is used for determinig the interaction between the spherical host and the source [21]–[23].

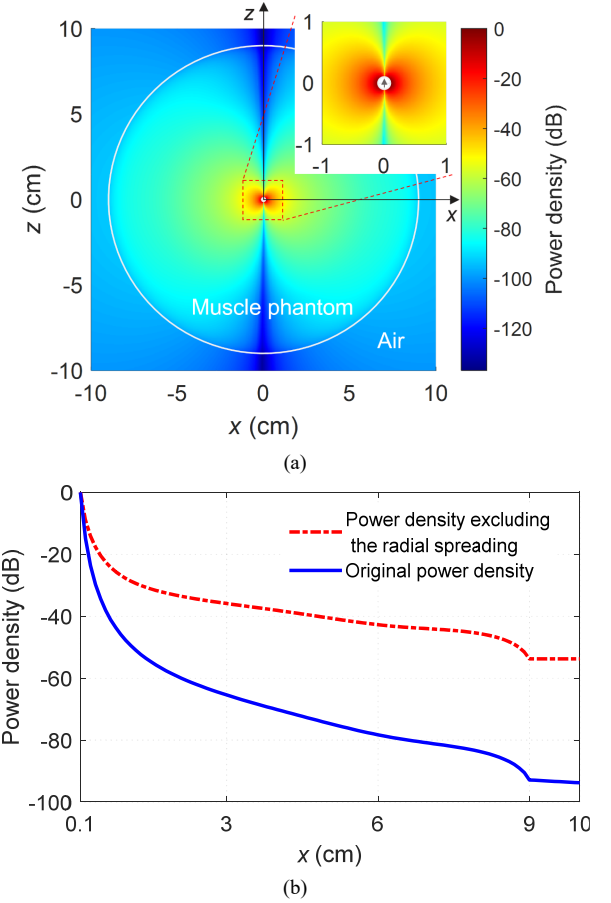

Fig. 2. An electric dipole ($r_{\mathrm{impl}}$=1 mm, at the origin) implanted in a spherical muscle phantom ($r_{\mathrm{body}}$=9 cm) at 403.5 MHz: (a) normalized power density distribution on the $x$-$z$ plane via the SWE method; (b) normalized power density reaching out of the spherical phantom along the $x$-axis via the SWE method, where the radial spreading effect is excluded for the red dashed line.

Consider the example of an electric dipole source implanted at the centre of a spherical muscle phantom with $r_{\mathrm{impl}} = 1$ mm, $r_{\mathrm{body}} = 9$ cm and the operating frequency is 403.5 MHz. Using the SWE method, we can accurately obtain the EM fields inside and outside the phantom, and the power density through the calculation of the Poynting's vector. In Fig. 2 (a), the power density distribution is displayed on the $x$-$z$ plane (as a reference value we took the power density value just outside the lossless sphere containing the implanted antenna). Owing to the spherical symmetry of the considered model, the distribution of the power density is symmetric around the $z$-axis, and the values decay significantly through the muscle phantom before reaching the phantom–air interface. If we look at the distribution of the power density in the maximum gain direction, e.g., the positive $x$-direction, we obtain the power density as a function of the radial distance, as illustrated by the blue solid curve in Fig. 2 (b). The decay of the power density here also includes the radial spreading effect, because the area of the sphere where the power density is collected increases quadratically with the increase of the radial distance. To eliminate the effect of the radial spreading and concentrate on the losses due to the medium, we multiply the power density value by the square of the radial distance. It can be observed that for this centred source model, the power density thus normalized remains constant after reaching the air, and indeed has the same value as the radiation efficiency in this case, representing the losses in the medium.

## III. Approximations to the Reactive Near-Field Losses

As shown in previous works [14] and [16], it is possible to express the maximum power density reaching the free space using expressions for different sources of losses and corresponding efficiency terms (denoted with $e$) [14]:

$$\text{Re}\left\{S_{\substack{\text{reaching} \\ \text{free space}}}\right\} = \text{Re}\left\{S_{\substack{\text{entering} \\ \text{the body}}}\right\} \cdot \frac{r_{\text{impl}}^2}{d^2} \cdot e_{\text{total}}$$

$$= \text{Re}\left\{S_{\substack{\text{entering} \\ \text{the body}}}\right\} \cdot \frac{r_{\text{impl}}^2}{d^2} \cdot e_{\substack{\text{losses in the} \\ \text{reactive} \\ \text{near-field}}} \cdot e_{\substack{\text{propagating} \\ \text{field absorption} \\ \text{losses}}} \cdot e_{\substack{\text{losses due to} \\ \text{reflections}}} \quad (2)$$

The power density entering the body, $\text{Re}\left\{S_{\substack{\text{entering} \\ \text{the body}}}\right\}$, is related to the maximum value of the real part of the Poynting's vector component normal to the surface of the implanted antenna, i.e., just inside the lossy body. In the same way, the power density reaching the free space is defined and denoted as $\text{Re}\left\{S_{\substack{\text{reaching} \\ \text{free space}}}\right\}$. Note that the factor $r_{\text{impl}}^2/d^2$ excludes the effect of the radial spreading of power density, as explained above. In (2), the total efficiency of the maximum power density is denoted as $e_{\text{total}}$, and the influence of the lossy body is expressed by three efficiency terms. The first term $e_{\substack{\text{losses in the} \\ \text{reactive} \\ \text{near-field}}}$ accounts for the losses due to the coupling of the implantable antenna's near field with the lossy medium. The second term, $e_{\substack{\text{propagating} \\ \text{field absorption} \\ \text{losses}}}$, represents the losses due to the propagating field and decays as $e^{-2k''(d-r_{\text{impl}})}$ ( $k''$ is the linear attenuation coefficient in the host body). Finally, the third one $e_{\substack{\text{losses due to} \\ \text{reflections}}}$ describes the losses at the phantom–air interface [14].

In this paper we will focus on the expressions for losses due to the reactive near field coupling only. To distinguish $e_{\substack{\text{losses in the} \\ \text{reactive} \\ \text{near-field}}}$ under different conditions, it is abbreviated in the rest of the paper as $e_{\text{NF, implantation depth}}^{\text{source type}}$. Here, the source type can be "TM" for electric dipole or "TE" for magnetic dipole; the implantation depth can be "deep" when $d > \lambda$ ($\lambda$ is the wavelength in the body tissue) or "shallow" when $d \leq \lambda$. As discussed in [14], the lowest near-field losses are achieved if the implanted source excite only the dominant spherical mode $n = 1$. On this basis, we first consider an electric dipole within a lossless capsule, which is placed in a homogenous lossy medium. The complex wave number of the lossy medium is $k = k' - ik''$. The radial component of the Poynting's vector power density excited by the source located in the origin of the spherical coordinate system becomes

$$\text{Re}\left\{S_r\right\} = \text{Re}\left\{E_\theta \cdot H_\varphi^*\right\}$$

$$= \sum_{n,m} \frac{C_{mn}}{r^2} \left[\frac{d}{d\theta} P_n^{|m|}(\cos\theta)\right]^2 \text{Re}\left[i\eta |k|^2 \cdot \hat{H}_n'^{(2)}(kr) \cdot \hat{H}_n^{(2)*}(kr)\right]$$

$$\stackrel{n=1}{=} C_{01} \frac{\sin^2\theta}{r^2} \text{Re}\left\{\eta\left[|k|^2 + 2k''r^{-1} + \left(1 - \frac{k^*}{k}\right)r^{-2} - ik^{-1}r^{-3}\right]\right\} e^{-2k''r}, \quad (3)$$

where $C_{mn}$ are constants related to the corresponding spherical modal coefficients, and in particular constant $C_{01}$ is determined by the power entering the lossy medium. Note that in the polynomial inside the square brackets, the "near-field part" and the "far-field part" can be distinguished according to the knowledge of the radiative EM fields of the electric dipole in free space. Furthermore, the last term $e^{-2k''r}$ is the one related to the losses due to the propagating field absorption.

If we suppose that the source is deeply implanted at the centre of the spherical body phantom with the implantation depth $d > \lambda$, i.e., the phantom–air interface is in the propagating field region within the body tissue, the losses caused by the reactive near field can be approximated as

$$e_{\text{NF,deep}}^{\text{TM}} = \frac{e^{2k''(r_{\text{far}}-r_{\text{impl}})}\text{Re}\left[i\eta \cdot \hat{H}_n'^{(2)}(kr_{\text{far}}) \cdot \hat{H}_n^{(2)*}(kr_{\text{far}})\right]}{\text{Re}\left[i\eta \cdot \hat{H}_n'^{(2)}(kr_{\text{impl}}) \cdot \hat{H}_n^{(2)*}(kr_{\text{impl}})\right]}$$

$$\stackrel{n=1}{=} \frac{|k|^2 \text{Re}(\eta)}{\text{Re}\left[\eta\left[|k|^2 + 2k''r_{\text{impl}}^{-1} + \left(1 - k^*/k\right)r_{\text{impl}}^{-2} - ik^{-1}r_{\text{impl}}^{-3}\right]\right]}. \quad (4)$$

This approximate expression was first introduced in [14], assuming that only the "propagating field components" can pass through the lossy medium and eventually become radiative fields. In other words, it is assumed that the antenna is implanted deep enough in the phantom to ensure that the near fields do not reach free space. To focus on the most dominant terms, the expression (4) can be further approximated as

$$e_{\text{NF,deep}}^{\text{TM}} \approx \frac{|k|^2 \text{Re}(\eta)}{\text{Re}\left[\eta\left(|k|^2 - ik^{-1}r_{\text{impl}}^{-3}\right)\right]} \approx \frac{|k|^2 \text{Re}(\eta)}{\text{Im}\left(\eta k^{-1}r_{\text{impl}}^{-3}\right)}. \quad (5)$$

However, for many implantation scenarios, the depth of the antenna is not large enough to ensure near fields are completely attenuated before reaching free space, especially when the frequency is relatively low as for instance in the MedRadio band. Indeed, the near-field region in the body depends on the electrical distance to the source (i.e., with respect to the wavelength $\lambda$ in the lossy tissue). It is worth noting that in the expression of total radiated power, the presence of near-field components at the phantom–air interface shows that part of the non-radiative reactive near field is transformed into radiated fields outside the lossy body. Thus, for cases with relatively shallow implantation depth ($d \leq \lambda$), the approximated near-field losses can be reduced taking into account the corresponding radiated power. Therefore, considering the implantation depth $d$, an accurate approximate expression for shallow implants can be derived as

$$
\begin{aligned}
e_{\text{NF,shallow}}^{\text{TM}} &= \frac{e^{2k''(d-r_{\text{impl}})} \operatorname{Re}\left[ i\eta \cdot \hat{H}_n'^{(2)}(kd) \cdot \hat{H}_n^{(2)*}(kd) \right]}{\operatorname{Re}\left[ i\eta \cdot \hat{H}_n'^{(2)}(kr_{\text{impl}}) \cdot \hat{H}_n^{(2)*}(kr_{\text{impl}}) \right]} \\
&\stackrel{n=1}{=} \frac{\operatorname{Re}\left\{ \eta \left[ |k|^2 + 2k''d^{-1} + \left(1 - k^*/k\right)d^{-2} - ik^{-1}d^{-3} \right] \right\}}{\operatorname{Re}\left\{ \eta \left[ |k|^2 + 2k''r_{\text{impl}}^{-1} + \left(1 - k^*/k\right)r_{\text{impl}}^{-2} - ik^{-1}r_{\text{impl}}^{-3} \right] \right\}}.
\end{aligned}
\tag{6}
$$

In fact, according to the derivation procedure, this expression can also be applied to the cases with deep implantation depth. To focus on the most dominant terms in (6), it can be further approximated as

$$
e_{\text{NF,shallow}}^{\text{TM}} \approx \frac{\operatorname{Re}\left[ \eta \left( |k|^2 - ik^{-1}d^{-3} \right) \right]}{\operatorname{Re}\left[ \eta \left( |k|^2 - ik^{-1}r_{\text{impl}}^{-3} \right) \right]}.
\tag{7}
$$

With the same analysis procedure, for the magnetic dipole case with the implantation depth $d > \lambda$, the expression of near-field losses is

$$
\begin{aligned}
e_{\text{NF,deep}}^{\text{TE}} &= \frac{e^{2k''(r_{\text{far}} - r_{\text{impl}})} \operatorname{Re}\left[ i\eta^{-1} \cdot \hat{H}_n'^{(2)}(kr_{\text{far}}) \cdot \hat{H}_n'^{(2)*}(kr_{\text{far}}) \right]}{\operatorname{Re}\left[ i\eta^{-1} \cdot \hat{H}_n'^{(2)}(kr_{\text{impl}}) \cdot \hat{H}_n'^{(2)*}(kr_{\text{impl}}) \right]} \\
&\stackrel{n=1}{=} \frac{|k|^2 \operatorname{Re}\left( \eta^{-1} \right)}{\operatorname{Re}\left\{ \eta^{-1} \left[ |k|^2 + 2k''r_{\text{impl}}^{-1} + \left(1 - k^*/k\right)r_{\text{impl}}^{-2} - ik^{-1}r_{\text{impl}}^{-3} \right] \right\}} \\
&= \frac{|k|^2 \operatorname{Re}\left( \eta^{-1} \right)}{\operatorname{Re}\left[ \eta^{-1} \left( |k|^2 + 2k''r_{\text{impl}}^{-1} \right) \right]} \approx \frac{|k|^2}{2k''r_{\text{impl}}^{-1}}.
\end{aligned}
\tag{8}
$$

For the magnetic dipole case with relatively shallow implantation depth ($d \leq \lambda$), the approximation expression of near-field losses can be improved as

$$
\begin{aligned}
e_{\text{NF,shallow}}^{\text{TE}} &= \frac{e^{2k''(d-r_{\text{impl}})} \operatorname{Re}\left[ i\eta^{-1} \cdot \hat{H}_n'^{(2)}(kd) \cdot \hat{H}_n^{(2)*}(kd) \right]}{\operatorname{Re}\left[ i\eta^{-1} \cdot \hat{H}_n'^{(2)}(kr_{\text{impl}}) \cdot \hat{H}_n^{(2)*}(kr_{\text{impl}}) \right]} \\
&\stackrel{n=1}{=} \frac{\operatorname{Re}\left\{ \eta^{-1} \left[ |k|^2 + 2k''d^{-1} + \left(1 - k^*/k\right)d^{-2} - ik^{-1}d^{-3} \right] \right\}}{\operatorname{Re}\left\{ \eta^{-1} \left[ |k|^2 + 2k''r_{\text{impl}}^{-1} + \left(1 - k^*/k\right)r_{\text{impl}}^{-2} - ik^{-1}r_{\text{impl}}^{-3} \right] \right\}} \\
&= \frac{|k|^2 + 2k''d^{-1}}{|k|^2 + 2k''r_{\text{impl}}^{-1}}.
\end{aligned}
\tag{9}
$$

It should be noted that in the final expression of (8) and (9), there are no high-order terms, which indicates that magnetic dipole has much less absorbed power in the near field. This is due to the fact that the near field of the magnetic dipole is essentially magnetic in nature, whereas the near field excited by electric dipoles is essitally electric in nature. The latter couples much more stronly with the dielctric losses of biological tissues.

Note also that the same approximate expressions for different types of efficiencies are used as well when calculting the upper bound of radiation efficiency, see Appendix for details.

## IV. Results and Applications of the Developed Expressions

### A. An Implanted Dipole Source within a Spherical Body Phantom

As a first example, we analyze a spherical body model, that is a dipole source implanted at the centre of a spherical body phantom (i.e., $d = r_{\text{body}}$), as illustrated in Fig. 1. The homogeneous phantom is composed of muscle, and the air sphere surrounding the source has a radius $r_{\text{impl}} = 1$ mm. Both the electric dipole and magnetic dipole cases are considered. To mainly focus on the near-field losses and mitigate the reflection caused by the phantom–air interface, the spherical phantom is placed into a lossless muscle medium (with permittivity equal to the real part of the muscle permittivity). The radiation efficiency, denoted as $e_{\text{rad}}$, can be accurately calculated using the SWE method. Moreover, owing to the spherical symmetry of this centred source model, the normalized power density reaching the free space (after excluding the radial spreading effect) has the same value as the radiation efficiency, as demonstrated in the Appendix. Therefore, the accuracy of the proposed approximation expressions can also be evaluated by considering the radiation efficiency $e_{\text{rad}}$. For the losses due to the propagating field $e_{\substack{\text{propagating} \\ \text{field absorption} \\ \text{losses}}}$, it is only a function of depth and medium properties. For the reflection losses $e_{\substack{\text{losses due to} \\ \text{reflections}}}$, in this case its value is close to 1 due to the use of the lossless muscle outside of the phantom. Specially, for the near-field losses part in the approximation $e_{\substack{\text{losses in the} \\ \text{reactive} \\ \text{near-field}}}$, the two approximate methods discussed in Section III are adopted in the following calculations.

As shown in Fig. 3, approximated results obtained using $e_{\text{NF,shallow}}$ always better match the rigorous results from the SWE method, especially when the electrical length of $d$ becomes small. To numerically compare the results obtained by different methods, we choose two commonly used operating frequencies for implanted antennas, 400 MHz [close to the medical implant communication service (MICS) band at 402 to 405 MHz] and 2.4 GHz [close to one of the industrial, scientific and medical (ISM) bands at 2.4 to 2.5 GHz] to give the corresponding radiation efficiency $e_{\text{total}}$, as shown in Table I. For cases with both type of sources, the deviation between the approximated results and the rigorous SWE results increases when the electrical length of the implantation depth $d$ becomes smaller. For the electric dipole cases, the difference between the two approximated curves is larger owing to the strong coupling to the dielectric losses in the near-field region. These results indicate that the newly proposed approximate expressions represent a better tool to accurately evaluate the radiation efficiency, especially for shallow implants when the near field leaks outside the lossy phantom.

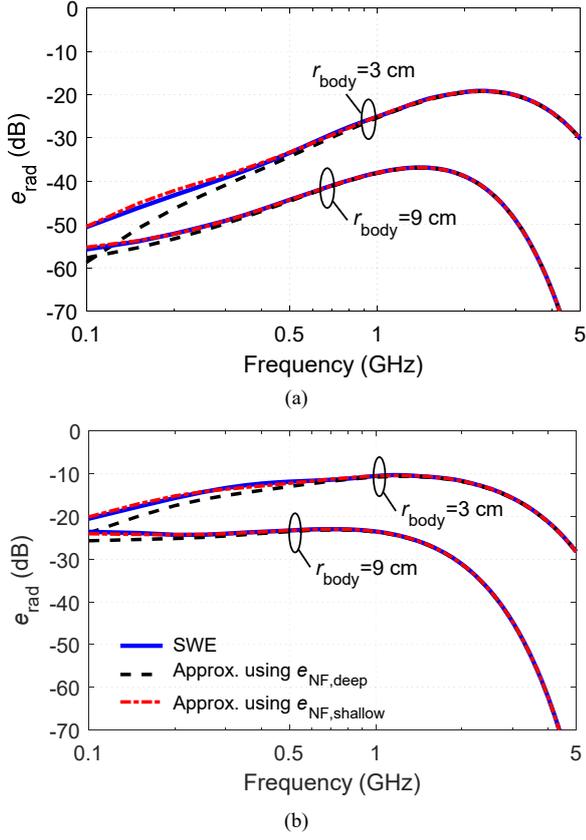

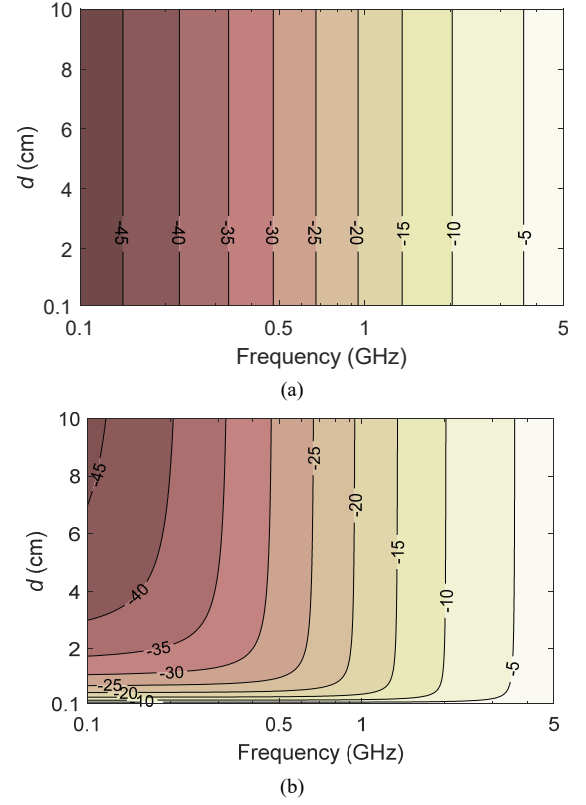

Fig. 3. Radiation efficiency $e_{rad}$ as a function of the frequency for excitations of (a) electric dipole source and (b) magnetic dipole source within the spherical body phantom.

Fig. 4. Distributions of the reactive near field losses as a function of the frequency and $d$ via the expressions of (a) $e_{NF,deep}^{TM}$ and (b) $e_{NF,shallow}^{TM}$.

TABLE I
RADIATION EFFICIENCY FOR IMPLANTED DIPOLE SOURCE WITHIN THE
SPHERICAL MODEL VIA DIFFERENT METHODS

| Source | Frequency | $d$ | $e_{rad}$ | | |
|---|---|---|---|---|---|
| | | | SWE | Approx. using $e_{NF,deep}$ | Approx. using $e_{NF,shallow}$ |
| Electric dipole | 400 MHz | 3 cm | −36.09 dB | −37.22 dB | −35.75 dB |
| | | 9 cm | −46.55 dB | −46.94 dB | −46.53 dB |
| | 2.4 GHz | 3 cm | −19.17 dB | −19.27 dB | −19.17 dB |
| | | 9 cm | −42.10 dB | −42.13 dB | −42.10 dB |
| Magnetic dipole | 400 MHz | 3 cm | −12.27 dB | −13.87 dB | −12.84 dB |
| | | 9 cm | −23.57 dB | −23.97 dB | −23.60 dB |
| | 2.4 GHz | 3 cm | −13.07 dB | −13.15 dB | −13.06 dB |
| | | 9 cm | −35.99 dB | −36.02 dB | −35.99 dB |

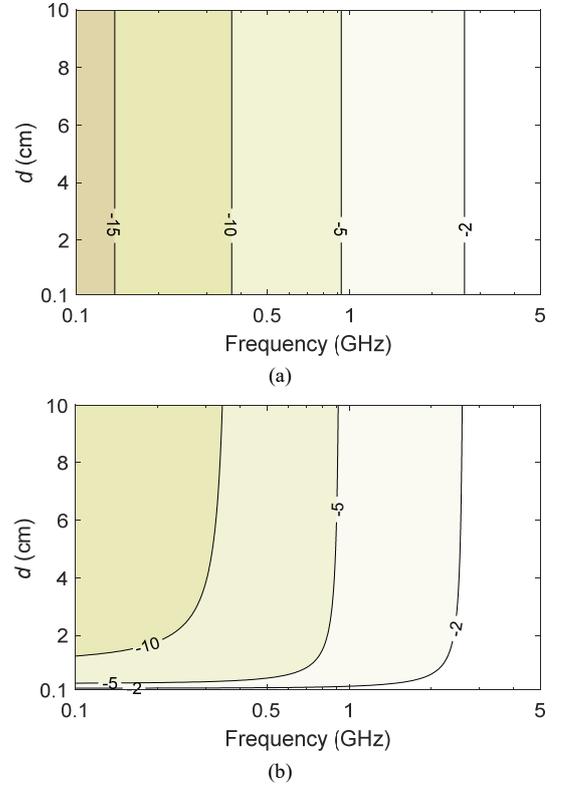

Fig. 5. Distributions of the reactive near field losses as a function of the frequency and $d$ via the expressions of (a) $e_{NF,deep}^{TE}$ and (b) $e_{NF,shallow}^{TE}$.

In Fig. 4 and 5, using the approximation methods, the reactive near-field losses can be plotted as a function of the operating frequency and the implantation depth $d$ for implanted electric and magnetic dipole with $r_{impl} = 1$ mm, respectively. For the results calculated using $e_{NF,deep}$, i.e. Fig. 4(a) and Fig. 5(a), the near-field loss does not vary with $d$. However, for the results calculated using $e_{NF,shallow}$, i.e. Fig. 4(b) and Fig. 5(b), the electrical length of $d$ comes into play. It can be seen that the near-field losses become a constant value at a specific frequency when $d$ is electrically large enough in the lossy medium; when this is not the case, especially for shallow implants, $d$ needs to be taken into account. Electric dipole sources have far more losses in the reactive near field than magnetic dipole sources, and the difference exceeds 10 dB when the frequency is below 1 GHz.

### B. An Implanted Dipole Source within a Planar Body Phantom

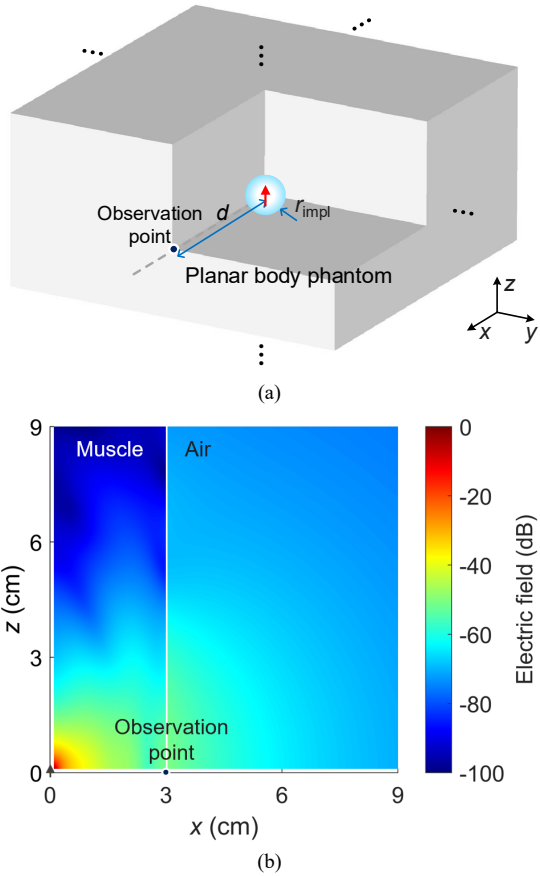

(a)

(b)

Fig. 6. (a) View of the planar body model with an elementary dipole source implanted with the deepth of $d$. (b) Normalized electric field distribution of an electric dipole (at the origin) implanted at the depth of 3 cm in a planar muscle phantom at 2.45 GHz via the Green's fuction method.

Although derived from the symmetrical spherical body model, the proposed approximate expressions indeed have a much broad range for applications. The most far away phantom model is a planar body model, i.e., the body phantom is represented by an infinite large lossy medium. Thus, the second example studies such a practical scenario for implants within an electrical large body, i.e., a planar body phantom with an antenna implanted at a depth of $d$, as shown in Fig. 6(a). Again, the antenna is modeled as a small air sphere with $r_{impl} = 1$ mm containing an elementary dipole source oriented in the $z$-direction, i.e. parallel to the closest phantom–air interface. The phantom is composed of muscle, and the outer space is filled with air. The EM fields and the power density inside and outside the phantom can be accurately calculated by the Green's function for multilayered planar media [24]–[27]. For example, for an electric dipole implanted at a depth of 3 cm operating at 2.45 GHz, the distribution of the normalized electric field on the $z$-$x$ plane is shown in Fig. 6, where the source is placed at the origin of the coordinate. At the point on the phantom–air interface closest to the source, i.e. the same location of the observation point in Fig. 6(a), the maximum electric field intensity reaching free space can be found.

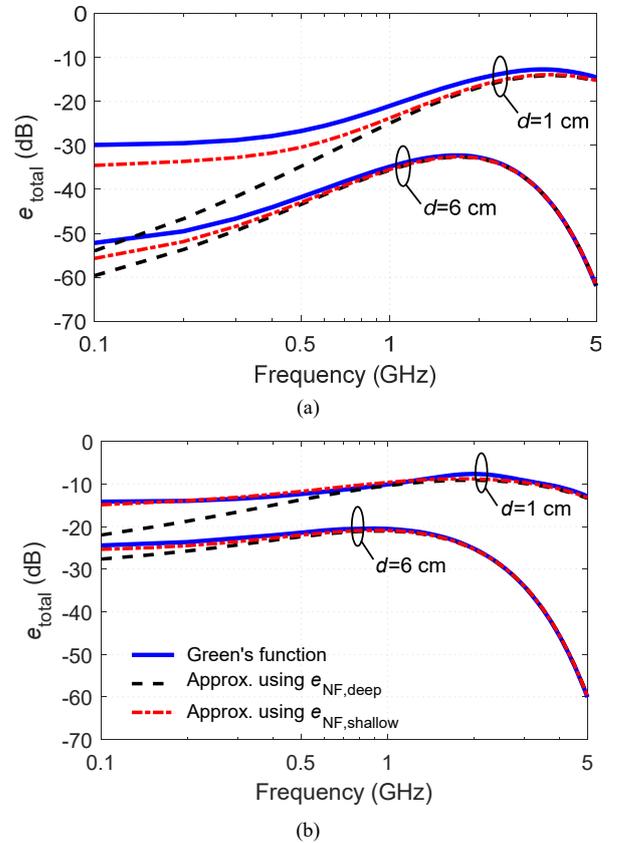

(a)

(b)

Fig. 7. Maximum power density as a function of the frequency for excitations of (a) electric dipole source and (b) magnetic dipole source within the planar body phantom.

With the approximate expressions of losses given in previous section, the efficiency of the maximum power density $e_{total}$ can also be evaluated analytically. Both approaches are compared in Fig. 7, again for an electric and for a magnetic source. We see that the values obtained using the analytical approximation for the near field efficiency presented in this paper, $e_{NF,shallow}$, agree well with the result obtained using the Green's function approach and additionally provides more accurate results than the approximation not taking into account the implantation depth, $e_{NF,deep}$. More explicitly, the numerical results obtained by different methods are summarized in Table II at 400 MHz and 2.45 GHz, respectively. This indicates that the proposed approximate expressions can be used to evaluate the maximum power density both in the case of shallow and deep implants independently on the shape of the body phantom.

| Source | Frequency | $d$ | $e_{total}$ | | |
|---|---|---|---|---|---|
| | | | Green's function | Approx. using $e_{NF,deep}$ | Approx. using $e_{NF,shallow}$ |
| Electric dipole | 400 MHz | 1 cm | –27.84 dB | –37.87 dB | –31.70 dB |
| | | 6 cm | –44.06 dB | –46.13 dB | –45.49 dB |
| | 2.4 GHz | 1 cm | –13.64 dB | –15.41 dB | –15.10 dB |
| | | 6 cm | –34.30 dB | –34.46 dB | –34.32 dB |
| Magnetic dipole | 400 MHz | 1 cm | –12.93 dB | –14.96 dB | –12.40 dB |
| | | 6 cm | –22.03 dB | –23.21 dB | –22.67 dB |
| | 2.4 GHz | 1 cm | –8.035 dB | –9.297 dB | –9.028 dB |
| | | 6 cm | –28.27 dB | –28.35 dB | –28.31 dB |

## C. A Realistic Implantable Capsule-Shaped Antenna within a Cube-Shaped Body Phantom

To validate the usefulness of the proposed method in practical applications, a realistic capsule-shaped implanted antenna is investigated. As shown in Fig. 8, for a cube-shaped body phantom consisting of muscle tissue with a side length $L_{body}$ = 15 cm, a capsule-shaped antenna is implanted at a depth $d$ = 3 cm below the centre of its left front side face. This implanted antenna is an electrically small dipole antenna consisting of two conductive cylinders with the radius of $r_{dip}$ = 0.06 mm and an overall length of $L_{dip}$ = 2 mm. To represent a general capsule or pill shape encapsulation, the antenna is encapsulated in a lossless air capsule structure, where the main body is a cylinder of radius $r_{encap}$ = 1 mm and terminated by two hemispherical ends with the same radius. According to the analysis of implanted capsule by Nikolayev et al. [13], the effective radius of the encapsulation $r_{impl}$ can be expressed by the circumradius of its main body, i.e., $r_{impl} = \sqrt{L_{dip}^2/4 + r_{encap}^2} \approx 1.41$ mm. In Fig. 8, the capsule antenna is oriented in the $z$-direction (i.e., parallel to the nearest phantom–air interface) to maximize power density in the positive $x$-direction (i.e., perpendicular to the interface). In order to achieve the shortest wireless link from the implanted antenna to the outside space, an observation point is placed at the phantom–air interface in the same direction.

This implanted antenna case is simulated by a finite element method (FEM) solver (Ansys Electronics Desktop 2020, HFSS). To avoid the reflection loss at the feed port, the impedance of the excitation lumped port is conjugated matched with the port impedance of the antenna itself. The operating frequency is chosen as 2.45 GHz.

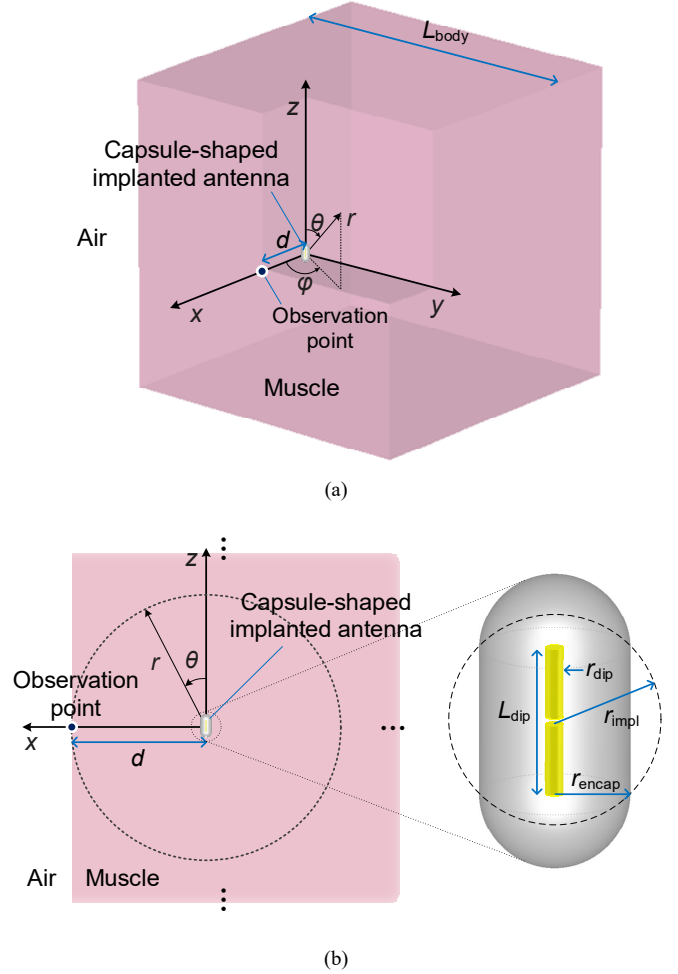

Fig. 8. (a) 3D View and (b) 2D side view on the $x$-$z$ plane of a realistic capsule-shaped antenna implanted at the depth of 3 cm within a cube-shaped muscle phantom, where the operating frequency of the antenna is 2.45 GHz.

Table III gives the simulated near-field gain and the efficiency of the maximum power density $e_{total}$ (calculated using approximate expressions) at the observation point. According to the definition of $e_{total}$, it can be interpreted as the near-field gain at the observation point with respect to a reference short dipole. Specifically, in order to convert $e_{total}$ to a gain value in dBi, its value needs to be enlarged by 1.76 dB, which is the directivity a short dipole. Ultimately, after conversion, the approximate result and the numerical solution for near-field gain in dBi at the observation point are within 1 dB of deviation. This indicates that the approximate expressions can be effectively used to estimate the near-field gain of the implanted antenna to an on-body antenna, which is also regarded as the upper limit of the gain reaching to the free space.

In Fig. 9 the power density, as a function of the radial distance in the positive $x$-direction, is calculated using both the general FEM tool and the approximate expressions. Again, the approximate expressions are derived by replacing the implanted antenna with an electric dipole encapsulated in a spherical air bubble with $r_{impl}$ = 1.41 mm, and implanted at the centre of a spherical muscle phantom with the radius $r_{body}$ = $d$. More specifically, we apply the two efficiency terms related to near-field losses and the propagating field losses within the body phantom (by letting $d = x$), and the term related to reflection losses is taken into account at the phantom–air interface. In both cases we use the value at $x = r_{impl}$ as the reference value.

TABLE III

Comparison of FEM Simulation and Approximate Results of the Realistic Capsule-Shaped Antenna

| FEM simulation | Approximate expressions | |
|---|---|---|
| Near-field gain at the observation point | $e_{total}$: Approx. using $e_{NF,shallow}$ | Conversion of $e_{total}$ to gain |
| –17.17 dBi | –19.78 dB | –18.02 dBi |

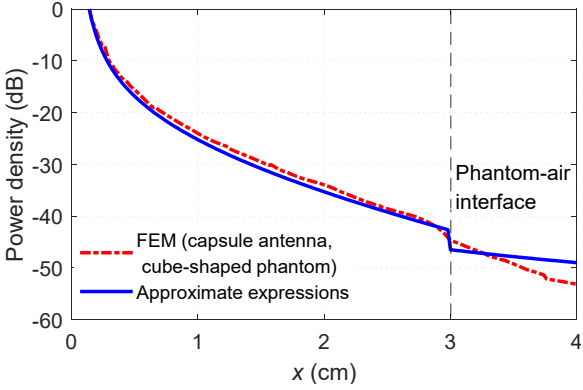

Fig. 9. Normalized power density reaching out of the body phantom along the $x$-axis via the FEM simulation and the proposed approximate expressions.

Within the muscle tissue, i.e., $x \leq 3$ cm, specifically in the near-field region, the power density curves of both cases are almost overlapping, which well validates the proposed approximate expressions to the reactive near-field losses. For the power density propagating outside the body phantom, the results for the spherical model are only decayed with the radial spreading due to the spherical symmetry of the considered model. However, the cube-shaped phantom has a plane aperture on the interface. This implies that outside the body we are in the near-field of the radiation aperture, so the same $1/r^2$ dependency will be obtained but no longer originating from the implanted source. Thus, the power density decays faster with radial distance compared to that of the spherical phantom.

## V. Conclusions

The coupling of the antenna reactive near field and the lossy body is a crucial part of implanted antennas regarding radiation efficiency, as it is the only loss contribution that can be improved through clever antenna design. In this work, we discussed several simplified body models for estimating the losses of antennas implanted into lossy biological tissue, specifically the losses due to the near-field coupling.

Based on previous works using a spherical wave decomposition to analyze the radiation of implanted antennas, accurate approximate expressions for the near-field losses are proposed in this paper, in which the implantation depth (in terms of the wavelength in the biological tissue) is taken into account. The obtained approximations are thus useful for both deep and shallow implantation scenarios. Although these expressions are derived using the canonical spherical body model with dominant spherical mode excitation (known yielding the lowest achievable near-field losses among the various excitation modes), they can also be used in practical scenarios with different body shapes. Indeed, the validity of the proposed expressions has been demonstrated for two very different body phantoms – a spherical phantom and a planar phantom. Furthermore, results for a real antenna in a capsule implanted in a cube-shaped body model are presented, and FEM simulations are compared with our approximations for the losses, showing a good agreement. These examples have demonstrated the usefulness of the approximate method in the design stage of an antenna, where the method can deliver quick access to link budgets and greatly assist in making the initial choices on antenna requirements.

## Appendix

For a centred source within a spherical body phantom, owing to the spherical symmetry of the considered model, the three efficiency terms in (2) can also be used to evaluate the losses of total radiated power due to the lossy body phantom. Specifically, the radiation efficiency $e_{rad}$ excluding the mismatching losses can be expressed in the following form:

$$
\begin{aligned}
P_{\substack{reaching \\ free space}} &= P_{\substack{entering \\ the body}} \cdot e_{rad} \\
&= P_{\substack{entering \\ the body}} \cdot e_{\substack{losses in the \\ reactive \\ near-field}} \cdot e_{\substack{propagating \\ field absorption \\ losses}} \cdot e_{\substack{losses due to \\ reflections}},
\end{aligned}
\tag{a1}
$$

where $P_{\substack{entering \\ the body}}$ is the total power entered into the lossy body phantom, $P_{\substack{reaching \\ free space}}$ is the total radiated power reaching free space, and the three efficiency terms have the same definition of the losses.

As analyzed in [14], for an electric dipole immersed in a lossy medium with the wave number of $k = k' - ik''$, the outward-directed radiated power $P_{total}$ over a sphere of radius of $r$ can be integrated as

$$
\begin{aligned}
P_{total} &= \int_0^{2\pi} \int_0^{\pi} \mathrm{Re}\left(E_\theta \cdot H_\varphi^*\right) \cdot r^2 \sin\theta \, d\theta \, d\varphi \\
&\stackrel{n=1}{=} C \frac{8\pi}{3} \mathrm{Re}\left\{\eta \left[ |k|^2 + 2k''r^{-1} + \left(1 - \frac{k^*}{k}\right)r^{-2} - ik^{-1}r^{-3} \right] \right\} e^{-2k''r}.
\end{aligned}
\tag{a2}
$$

Looking back at the expressions for the radial component of the Poynting's vector power density, i.e.,

$$
\begin{aligned}
\mathrm{Re}\{S_r\} &= \mathrm{Re}\left\{E_\theta \cdot H_\varphi^*\right\} \\
&\stackrel{n=1}{=} C \frac{\sin^2\theta}{r^2} \mathrm{Re}\left\{\eta \left[ |k|^2 + 2k''r^{-1} + \left(1 - \frac{k^*}{k}\right)r^{-2} - ik^{-1}r^{-3} \right] \right\} e^{-2k''r},
\end{aligned}
\tag{a3}
$$

one can see that the difference in these two equations is only in the factor $\sin^2\theta/r^2$ describing the effects of angular dependency and radial spreading of power density. This indicates that the same expressions for the loss approximation $e_{\text{losses in the reactive near-field}}$ and $e_{\substack{\text{propagating} \\ \text{field absorption} \\ \text{losses}}}$ are used when calculating the maximum power density (after excluding the radial spreading effect) and radiation efficiency once the source is implanted at the centre of a spherical body phantom. As for $e_{\substack{\text{losses due to} \\ \text{reflections}}}$, as given in [14], it can be well estimated using the spherical wave impedance of the fundamental mode for a centred source model, both in terms of power density and total power. The above conclusions also apply to the cases of magnetic dipole. Note that even if these expressions are developed for an antenna implanted in the center of a sphere radiating only the dominant TE or TM mode, they actually represent the upper bound for the total radiated power.

## Acknowledgment

The authors would like to thank Hannes Bartle and Ismael Vico Trivino for their help with simulations.

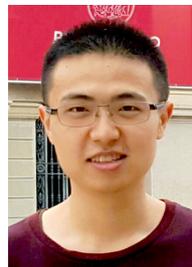
**Mingxiang Gao** was born in Taiyuan, China. He received the double B.Sc. degree in electrical engineering and business administration and the M.Sc. degree in electrical engineering from Xi'an Jiaotong University, Xi'an, China, in 2016 and 2019, respectively, and the M.Sc. degree (summa cum laude) in electrical engineering from the Politecnico di Milano, Milan, Italy, in 2019. He is currently working towards the Ph.D. degree at the Ecole Polytechnique Federale de Lausanne, Lausanne, Switzerland. His current research interests include theoretical analysis and design of implantable antennas and wireless techniques for biomedical electronics.

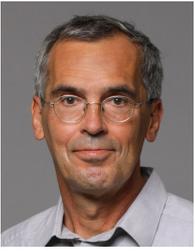

**Zvonimir Sipus (Senior Member, IEEE)** received the B.Sc. and M.Sc. degrees in electrical engineering from the University of Zagreb, Zagreb, in 1988 and 1991, respectively, and the Ph.D. degree in electrical engineering from the Chalmers University of Technology, Gothenburg, Sweden, in 1997. From 1988 to 1994, he was with the Rudjer Boskovic Institute, Zagreb, as a Research Assistant, where he was involved in the development of detectors for explosive gasses. In 1994, he joined the Antenna Group, Chalmers University of Technology, where he was involved in research projects concerning conformal antennas and soft and hard surfaces. In 1997, he joined the Faculty of Electrical Engineering and Computing, University of Zagreb, where he is currently a Professor. From 1999 to 2005, he was also an Adjunct Researcher with the Department of Electromagnetics, Chalmers University of Technology. Since 2006, he has been involved in teaching with the European School of Antennas. From 2008 to 2012 and from 2014 to 2018, he has been the Head of the Department of Wireless Communications. His current research interests include the analysis and design of electromagnetic structures with application to antennas, microwaves, and optical communication and sensor systems.

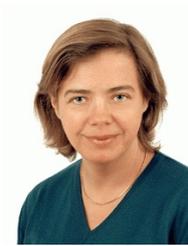

**Anja K. Skrivervik** received the master's degree in electrical engineering and the Ph.D. degree from the Ecole Polytechnique Fédérale de Lausanne (EPFL), Lausanne, Switzerland, in 1986 and 1992, respectively. She was an Invited Research Fellow with the University of Rennes, Rennes, France, followed by two years in the industry. In 1996, she rejoined the EPFL as an Assistant Professor, where she is currently a Full Professor and also the Head of the Microwaves and Antennas Group. Her teaching activities include courses on microwaves and antennas and courses at Bachelor, Master, and Ph.D. levels. She has authored or coauthored more than 200 peer-reviewed scientific publications. Her current research interests include electrically small antennas, antennas in biological media, multifrequency and ultrawideband antennas, and numerical techniques for electromagnetics. She has been a Member of the Board of Directors of the European Association on Antennas and Propagation, since 2017, and is a Board Member of the European School on Antennas. She was the recipient of the Latsis Award. She is frequently requested to review research programs and centres in Europe. She was the Chairperson of the Swiss URSI until 2012. She was the Director of the EE section from 1996 to 2000, and is currently the director of the EE doctoral school at EPFL. She is very active in European collaboration and European projects.